\documentclass[%
 reprint,
groupedaddress,
 amsmath,amssymb,
 aps,
 showkeys,
prb,
floatfix,
]{revtex4-2}

\usepackage{graphicx}
\usepackage{dcolumn}
\usepackage{bm}
\usepackage{hyperref}
\hypersetup{colorlinks=true,citecolor=blue,linkcolor=red,urlcolor=blue}
\usepackage[mathlines]{lineno}
\usepackage{siunitx}
\usepackage{caption}
\usepackage{float}
\usepackage[english]{babel}

\newcommand{\cubcrys}{NaK\textsubscript{2}Sb}
\newcommand{\hexcrys}{Na\textsubscript{2}KSb} 
\newcommand{\dft}{density-functional theory} 
 
\newcommand{\mbpt}{many-body perturbation theory} 
 
\newcommand{\GW}{\ensuremath{GW}} 
\newcommand{\Gzero}{\ensuremath{G_0W_0}}

\DeclareSIUnit\angstrom{\text {Å}}
\begin{document}
\title{Electronic and optical properties of computationally predicted Na-K-Sb crystals}

\author{Chung Xu}
 \email{chung.ping.xu@uni-oldenburg.de}
\affiliation{%
 Carl von Ossietzky Universit\"at Oldenburg, Institute of Physics, 26129 Oldenburg,
Germany
}%

\author{Richard Schier}
 \email{richard.schier@uni-oldenburg.de}
\affiliation{%
 Carl von Ossietzky Universit\"at Oldenburg, Institute of Physics, 26129 Oldenburg,
Germany
}%

\author{Caterina Cocchi}
\email{caterina.cocchi@uni-oldenburg.de}
\affiliation{%
 Carl von Ossietzky Universit\"at Oldenburg, Institute of Physics, 26129 Oldenburg,
Germany
}%

\date{\today}

\begin{abstract}
Thanks to their favorable electronic and optical properties, sodium-potassium-antimonides are an emerging class of crystals used as photocathodes in particle accelerators. The persisting challenges related to the synthesis and characterization of these materials demand support from theory and make the study of computationally predicted polymorphs particularly relevant to identifying the structure and composition of the samples. Using first-principles methods based on \dft~and \mbpt, the electronic and optical properties of cubic \cubcrys~and hexagonal \hexcrys~are studied. Both systems, most commonly found in the hexagonal and cubic phase, respectively, exhibit an indirect fundamental gap that is energetically very close to the direct band gap at $\Gamma$ of magnitude $\SI{0.81}{\eV}$ for \cubcrys~and $\SI{0.70}{\eV}$ for \hexcrys. In the band structure of both materials, Sb $p$-states dominate the valence region with minor contributions from the alkali $p$-states, while the alkali $s$-states mainly contribute at lower energies. The optical spectra of both crystals are not subject to sizeable excitonic effects, except for a redshift of the excitation energies of the 50-100~meV and some redistribution of the oscillator strength beyond the lowest-energy peak in the near-infrared region. 
Our results indicate that computationally predicted cubic \cubcrys~and hexagonal \hexcrys~have favorable characteristics as photocathodes and, as such, their presence in polycrystalline samples is not detrimental for these applications.

\end{abstract}

\maketitle

\section{\label{sec:intro}Introduction}%
Multi-alkali antimonides (MAAs) have attracted considerable attention in the last couple of decades thanks to their potential use in several applications ranging from cryogenics to photomultiplier tubes \cite{BIALKALI_cryogenic, BIALKALI_hirano-2013, BIALKALI_MOTTA, BIALKALI_NAKAMURA2010276}. More recently, they have been successfully proposed as photocathode materials for particle accelerators~\cite{ACC_10.1063/1.2883909,Na2KSb_QE_experiment,PhysRevAccelBeams.21.113401,moha+23micromachines} due to their absorption of infrared radiation, whilst maintaining or improving on the low thermal emittance obtained by bi-alkali and mono-alkali antimonides \cite{EBEAM_10.1063/1.4945091,BIALKALI_MOTTA}. MAAs also have found popular use in free electron lasers and as a superior source of spin-polarized electrons compared to type III-IV (e.g., GaAs) photocathodes \cite{SPIN_SOURCE_PhysRevLett.129.166802,ADVANCES_ESOURCE+MUSUMECI2018209}, which are typically very sensitive to residual gases and consequently have a too short lifetime for these applications~\cite {GaAS_matejcek}. An efficient electron source requires a high quantum efficiency, which represents the amount of emitted photoelectrons in comparison to the amount of incident photons. MAAs offer advantages over popular metals such as copper and magnesium because the latter typically have a higher reflectivity and shallow escape depth \cite{Mg_wang:linac02-mo445,KONG1995272, Na2KSb_QE_experiment}. Additional studies have shown that MAAs have matching or superior quantum efficiency to popular semiconductor materials such as GaAs \cite{GaAS_matejcek, syms-1969, chanlek-2015, debur-2009} and ultrananocrystalline diamond \cite{diamond_chen-2020,EBEAM_10.1063/1.4945091}, especially in the infrared range, where inefficient energy conversion processes are minimized while quantum efficiency is maximized~\cite{ADVANCES_ESOURCE+MUSUMECI2018209}. 

Recent \textit{ab initio} works have shed light on the fundamental electronic, spectroscopic, and thermal properties of MAAs~\cite{guo14mre,Cs_pressure_KALARASSE20101732,Cocchi_2019,Cs_cocchi-2019,cocc20pssrrl,anto+20prb,amador,schi+22prm,schi+24ats,wu-gano23jmca,csk2sb_PBE,Santana-Andreo_2024}, providing invaluable support to earlier experimental findings indicating this material class and in particular cesium-potassium antimonides as promising electron sources due to their small band gaps, low electron affinity, and sensitivity to visible light~\cite{CsKSb_nathan-1967,ghosh-1978,sommer-1955}. Sodium-potassium antimonides have been recently increasing in popularity due to a larger-wavelength absorption onset~\cite{wavelength_onset10.1063/1.1663009, monaco2022review}. Being lighter than cesium, sodium offers opportunities for developing photocathodes with advantageous physical characteristics whilst maintaining superior electronic and optical properties. For example, sodium-based MAAs are thermodynamically stable above $\SI{400}{\kelvin}$ \cite{ERJAVEC1994617} in contrast to some Cs-K-Sb compounds \cite{Santana-Andreo_2024}.
Unfortunately, MAAs are challenging to produce even with modern deposition methods \cite{QE_NaKSB_10.1063/1.4922146, 10.1063/5.0053186, Xie_2017, ERJAVEC1994617, PhysRevSTAB.14.120101} as they need to be grown under ultra-high vacuum conditions to prevent contamination with atmospheric pollutants \cite{Growth_electronics9121991,Cultrera_growth}. In addition to this high sensitivity limiting the chances to conduct thorough experimental diagnostics on the samples, there are issues with developing stable and reproducible deposition recipes. These problems enhance the role of computational studies, further supplemented by high-throughput screening methods~\cite{bai+20jpcc,wang+20prm,sass-cocc22jcp,sass-cocc24npjcm} and machine learning algorithms~\cite{anto+21am}, to predict and characterize MAA samples, which are often non-stoichiometric and polycrystalline~\cite{NAKSBPHASES_MCCARROLL196030, Cs_nonstoi, CsK_amorphous}. 
Databases collecting experimentally known and computationally predicted materials play an important role in identifying structure and composition. Employing state-of-the-art methods based on density-functional theory (DFT) and many-body perturbation theory (MBPT)~\cite{cocc-sass21micromachines} enables gaining valuable insight into the electronic and optical properties of MAAs, which is essential to interpret their experimental signatures.  

In this \textit{ab initio} study based on DFT and MBPT, we investigate the electronic and optical properties of two computationally predicted phases of sodium-potassium antimonide, namely hexagonal \hexcrys~and cubic \cubcrys. The experimentally known structures are cubic and hexagonal crystals, respectively, and their \textit{ab initio} characterization at the same level of theory was done in previous work~\cite{amador}. Here, we aim to complement the existing knowledge~\cite{amador, wavelength_onset10.1063/1.1663009, Na2KSb_QE_experiment} by exploring alternative phases that could be present as metastable polymorphs in the samples. After the characterization of the electronic structure of the two materials, we calculate their optical properties in terms of the macroscopic dielectric function with and without excitonic interactions. With this analysis, we gain insight into the spectral features of the two crystals and the composition of their optical excitations, which is essential to understanding their photoemission processes and assessing their photocathode performance.
\section{\label{sec:theory}Theoretical Background}%
The results presented in this work are performed in the framework of DFT~\cite{DFT_PhysRev.136.B864} and MBPT~\cite{BSE_1RevModPhys.74.601}, including the $GW$ approximation \cite{Hedin1965:PhysRev.139.A796} and the solution of the Bethe–Salpeter equation (BSE)~\cite{BSE_PhysRev.84.1232}.
Ground-state calculations are performed by solving the Kohn-Sham (KS) equations~\cite{KS_PhysRev.140.A1133} of DFT,
\begin{equation}
\left[\hat{H}^\text{KS}(\mathbf{r}) - \epsilon_i^\text{KS}\right] \psi_i(\mathbf{r}) = 0,
\end{equation}
where $\epsilon_i$ and $\psi_i(\mathbf{r})$ are the energies and wave-functions of the KS electrons, respectively, with the KS Hamiltonian in atomic units,
%
\begin{equation}
 \hat{H}^\text{KS}(\mathbf{r}) = - \frac{1}{2}\nabla^2 + V^\text{KS}(\mathbf{r}),
    \label{eq:KS}
\end{equation}
including the kinetic energy term and the KS potential
\begin{equation}
V^\text{KS}(\mathbf{r}) = V^\text{ext}(\mathbf{r}) + V^\text{Hartree}(\mathbf{r}) + V^\text{XC}(\mathbf{r}),
\end{equation}
embedding the Coulomb attraction exerted by the nuclei ($V^\text{ext}$), the Hartree potential ($V^\text{Hartree}$), and the exchange-correlation potential ($V^\text{XC}$). 
Since the exact form of $V^\text{XC}$ is not known, the accuracy of the solutions of the KS equations depends on the approximations applied to this term~\cite{VXC_accuracy, VXC_accuracy_borlido-2020}. 

MBPT is the state-of-the-art method to describe the excited-state properties of solids~\cite{BSE_1RevModPhys.74.601}. The \GW~approximation~\cite{Hedin1965:PhysRev.139.A796} is applied on top of the solutions of the KS equations to account for the quasi-particle (QP) correction to the electronic states, including the band gap, through the so-called QP equation
\begin{align}\label{eqn:GWQP}
   \epsilon_{\mathbf{k}}^\text{QP} = \epsilon_{\mathbf{k}}^\text{KS} + Z_{\mathbf{k}} \left[ {\Re}~\Sigma_{\mathbf{k}}(\epsilon_{\mathbf{k}}) - V_{\mathbf{k}}^\text{XC} \right],
\end{align}
where $\epsilon_{\mathbf{k}}^\text{KS}$ are the KS energies obtained from Eq.~\eqref{eq:KS} and $\Sigma_{\mathbf{k}}$ is the frequency-dependent, non-local self-energy given by
\begin{align}\label{eqn:selfenergy} 
&\Sigma_{\mathbf{k}}(\mathbf{r},\mathbf{r}';\omega) \\ \nonumber
&= \frac{i}{2\pi} \int G_0(\mathbf{r},\mathbf{r}';\omega+\omega')~W_0(\mathbf{r},\mathbf{r}';\omega')~e^{i\omega'\eta}~d\omega'.
\end{align}
In Eq.~\eqref{eqn:selfenergy}, $G_0$ is the non-interacting single-particle Green's function obtained from the KS states and $W_0$ is the dynamically screened Coulomb potential calculated in the random-phase approximation (RPA); the term $\eta$ in the exponential factor prevents the divergence of the integral and the QP renormalization factor $Z_{\mathbf{k}}$ accounts for the energy-dependence of $\Sigma_{\mathbf{k}}$. The single-shot solution of Eq.~\eqref{eqn:GWQP} is known as the \Gzero~approach~\cite{Hybertsen:PhysRevB.34.5390}, which often provides results in good agreement with experiments~\cite{GW_experiment} despite its lack of self-consistency~\cite{GW_selfconsistency}.%

Optical excitations are computed from the solution of the BSE~\cite{BSE_PhysRev.84.1232} on top of the QP energies computed from \Gzero. The problem is transformed into the eigenvalue equation%
\begin{equation}\label{eqn:BSE}
\sum_{v^{\prime} c^{\prime} \mathbf{k}^{\prime}} H_{v c \mathbf{k}, v^{\prime} c^{\prime} \mathbf{k}^{\prime}}^{\mathrm{BSE}} A_{v^{\prime} c^{\prime} \mathbf{k}^{\prime}}^\lambda=E^\lambda A_{v c \mathbf{k}}^\lambda,
\end{equation}
ruled by the two-particle BSE Hamiltonian, which for spin-singlet systems reads: $\hat{H}^\text{BSE} = \hat{H}^{\rm diag}+2\hat{H} ^{\rm x} +\hat{H}^{\rm c}$. 
The diagonal term, $H^\text{diag}$, accounts for transitions between valence-conduction ($v \rightarrow c$) QP states with $\mathbf{k}=\mathbf{k'}$; the exchange term, $H^{\rm x}$, describes the repulsive exchange interaction between fermions of opposite sign (electron and holes); the direct term, $H^{\rm c}$, embeds the screened electron-hole Coulomb attraction. 
In addition to excitation energies $E^\lambda$, the solution of Eq.~\eqref{eqn:BSE} outputs the eigenvectors $A^\lambda$ describing the character and composition of the excitations. 
This information can be visualized on top of the QP band structure through the definition of the so-called \textit{exciton weights} for holes 
\begin{equation}
w^{\lambda}_{v\mathbf{k}} = \sum_c |A^{\lambda}_{vc\mathbf{k}}|^2
\label{eq:w_h}
\end{equation}
and electrons
\begin{equation}
 w^{\lambda}_{c\mathbf{k}} = \sum_v |A^{\lambda}_{vc\mathbf{k}}|^2.
\label{eq:w_e}
\end{equation}
In the optical limit, the imaginary part of the macroscopic dielectric tensor, 
\begin{align}\label{eqn:macrdielectric}
\Im \varepsilon_M=\frac{8 \pi^2}{\Omega} \sum_\lambda\left|\mathbf{t}^\lambda\right|^2 \delta\left(\omega-E^\lambda\right),
\end{align}
 yields the optical absorption spectrum, where the transition coefficients 
\begin{equation}\label{eqn:transitioncoeffs}
\mathbf{t}^\lambda=\sum_{v c \mathbf{k}} A_{v c \mathbf{k}}^\lambda \frac{\langle v \mathbf{k}|\hat{\mathbf{p}}| c \mathbf{k}\rangle}{\epsilon_{c \mathbf{k}}^{\rm QP}-\epsilon_{v \mathbf{k}}^{\rm QP}},
\end{equation}
include the unit cell volume $\Omega$ and the angular frequency of the incoming photon $\omega$ in addition to the BSE eigenfunctions $A_{v c \mathbf{k}}^\lambda$ and the momentum matrix elements.

\section{\label{sec:param}Computational Details}%
All calculations performed in this work are carried out with \texttt{exciting}, a full-potential all-electron code implementing the linearized augmented plane waves and local orbitals method for DFT and MBPT~\cite{exciting_Gulans_2014}.
In all simulations, the muffin-tin radii ($R_\text{MT}$) of both Na and K atoms are set to 2.0~bohr, while for Sb, $R_\text{MT} =$~2.2~bohr. A plane-wave basis-set cutoff of $R_{\rm MT}G_{\rm max} = 8.0$, where $G_\text{max}$ is the cutoff length for reciprocal lattice vectors, is taken for cubic \cubcrys, while for hexagonal \hexcrys~it is set to 8.5. The generalized-gradient approximation in the spin-unpolarized Perdew-Burke-Ernzerhof parameterization for solids (PBEsol)~\cite{GGA_PBE_SOL} is chosen to approximate $V^\text{XC}$. A $10\times 10\times 10$ ($8 \times 8 \times 4$) $\mathbf{k}$-mesh is adopted to sample the Brillouin zone of \cubcrys~(\hexcrys). Lattice optimization is performed by minimizing the crystal volume with the aid of a Birch-Murnaghan fit~\cite{Murnaghan_pressure, cubic_BM_PhysRev.71.809} with a threshold of 0.01~eV/\AA{} for the residual interatomic forces.

%

\Gzero~calculations~\cite{GW_PhysRevB.94.035118} are performed on top of the KS electronic structure using a $6 \times 6 \times 6$ \textbf{k}-mesh for cubic \cubcrys~and a $4 \times 4 \times 2$ \textbf{k}-mesh for hexagonal \hexcrys. Screening is evaluated in the RPA including 200 empty states for both crystals with the self-energy computed by analytic continuation. The BSE~\cite{Vorwerk_2019} is solved in the Tamm-Dancoff approximation. A $\Gamma$-shifted \textbf{k}-mesh with $8 \times 8 \times 8$~($8 \times 8 \times 4$) points is employed in the BSE calculations for cubic \cubcrys~(hexagonal \hexcrys). The screened Coulomb interaction is computed using the RPA and 100 empty states. A local field cutoff of 1.5~Ha is taken. The considered transition space includes 3 valence and 9 conduction bands for cubic \cubcrys, while for hexagonal \hexcrys, 6 valence states and 12 conduction states. A Lorentzian broadening of \SI{27}{\milli\eV} is used for plotting $\Im\varepsilon_M(\omega)$. The computational settings adopted here largely overlap with those employed in Ref.~\cite{amador}, thus facilitating a direct comparison between the two studies.

\section{\label{sec:results}Results}%
\subsection{\label{subsec:StructuralRes}Structural properties}
\begin{figure}[h]
\centering
\includegraphics[width=0.45\textwidth]{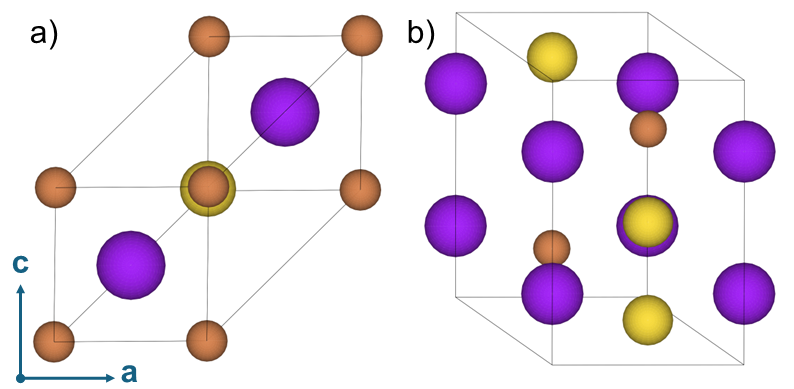}
\caption{Visualisation of a) cubic \cubcrys~and b) hexagonal \hexcrys~in their primitive unit cells. K atoms are depicted in purple, Sb atoms in bronze, and Na atoms in gold. Graphs produced with VESTA~\cite{Momma:db5098}.}
\label{fig:visualisation}
\end{figure}%

The structural parameters of the materials investigated in this work, namely cubic \cubcrys~and hexagonal \hexcrys, are available from the Open Quantum Materials Database \cite{OQMD}. \cubcrys~has a face-centred cubic Bravais lattice with Sb at Wyckoff position $(0,0,0)$, Na at $(1/2,1/2,1/2)$, and K at $\pm(1/4, 1/4, 1/4)$. \hexcrys~is a hexagonal crystal with the Sb atoms at $(2/3,1/3,3/4)$ and $(1/3,2/3,1/4)$, Na atoms at $(2/3,1/3,1/4 \pm 1/6)$, and $(1/3,2/3,3/4 \pm 1/6)$ and the K atoms at $(0, 0, \pm 1/4)$. The two structures are visualized in Fig.~\ref{fig:visualisation}. The lattice parameter of cubic \cubcrys~is $a = \SI{8.26}{\angstrom}$ which is larger than the one obtained at the same level of theory for the experimentally known cubic phase of \hexcrys~($a = \SI{7.74}{\angstrom}$)~\cite{amador}. This result is unsurprising considering the larger atomic radius of K compared to Na (Fig.~\ref{fig:visualisation}a). The in-plane and out-of-plane lattice parameters of hexagonal \hexcrys~are $a = \SI{5.77}{\angstrom}$ and $c = \SI{9.40}{\angstrom}$, respectively, resulting in a $c/a$ ratio of $1.63$. In this case, the experimentally known hexagonal \cubcrys~has a smaller in-plane lattice parameter $a = \SI{5.58}{\angstrom}$ but a larger out-of-plane one, $c = \SI{10.90}{\angstrom}$, giving rise to an overall larger $c/a$ ratio of 1.95~\cite{amador}. The reason for this behavior can be ascribed to the different compositions and positions of the constituting species in the unit cell (Fig.~\ref{fig:visualisation}b).

\subsection{\label{subsec:electronicres}Electronic structure}%

\begin{figure}[htb]
    \centering
\includegraphics[width=0.5\textwidth]{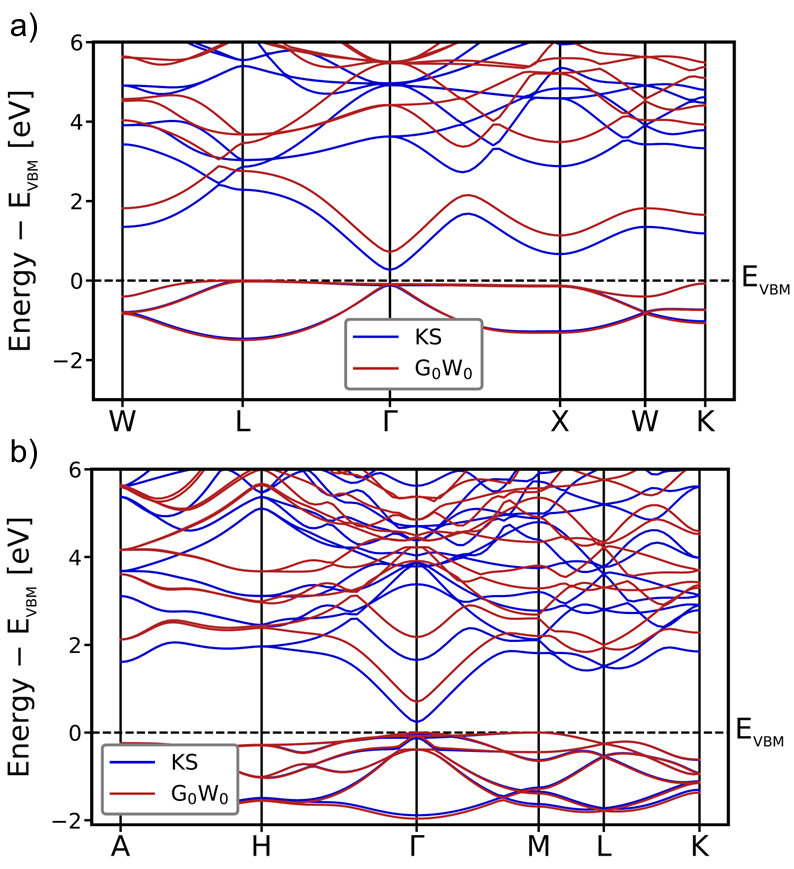}
    \caption{\label{fig:DFTBS} Electronic band structure of a) \cubcrys~and b) \hexcrys~computed from the KS scheme of DFT (blue) and \Gzero (red). 
    } 
\end{figure}


We begin our analysis of the electronic properties of \cubcrys~and \hexcrys~by inspecting their band structure shown in Fig.~\ref{fig:DFTBS}. Both crystals are characterized by an indirect band gap with the conduction band minimum (CBm) at $\Gamma$ and the valence band maximum (VBM) at the high-symmetry points L in cubic \cubcrys~(Fig.~\ref{fig:DFTBS}a) and M in hexagonal \hexcrys~(Fig.~\ref{fig:DFTBS}b). 
These characteristics contrast against the findings obtained at the same level of theory for the experimentally known cubic \hexcrys~and hexagonal \cubcrys, which indicate them as direct band-gap semiconductors~\cite{amador}. While DFT qualitatively captures the nature of the band gap, it significantly underestimates its magnitude compared to \Gzero, as expected. The self-energy contributes an increase of the fundamental gap by 0.43~eV: in \cubcrys, the band-gap changes from 0.38~eV (DFT) to 0.81~eV (\Gzero), and in \hexcrys, 0.27~eV to 0.70~eV.
In both crystals, the QP correction becomes more pronounced at higher energies within the conduction band. This trend is more evident in \cubcrys~due to the fewer bands in the considered energy range.
In the valence region, the lower-energy QP bands visualized in Fig.~\ref{fig:DFTBS} experience a slight downshift of a few tens of meV compared to their counterparts computed from DFT. In the experimentally-known phases, the self-energy renormalization gives rise to a 100~meV downshift~\cite{amador}.

\begin{figure*}[htb]
    \centering
      \includegraphics[width=\textwidth]{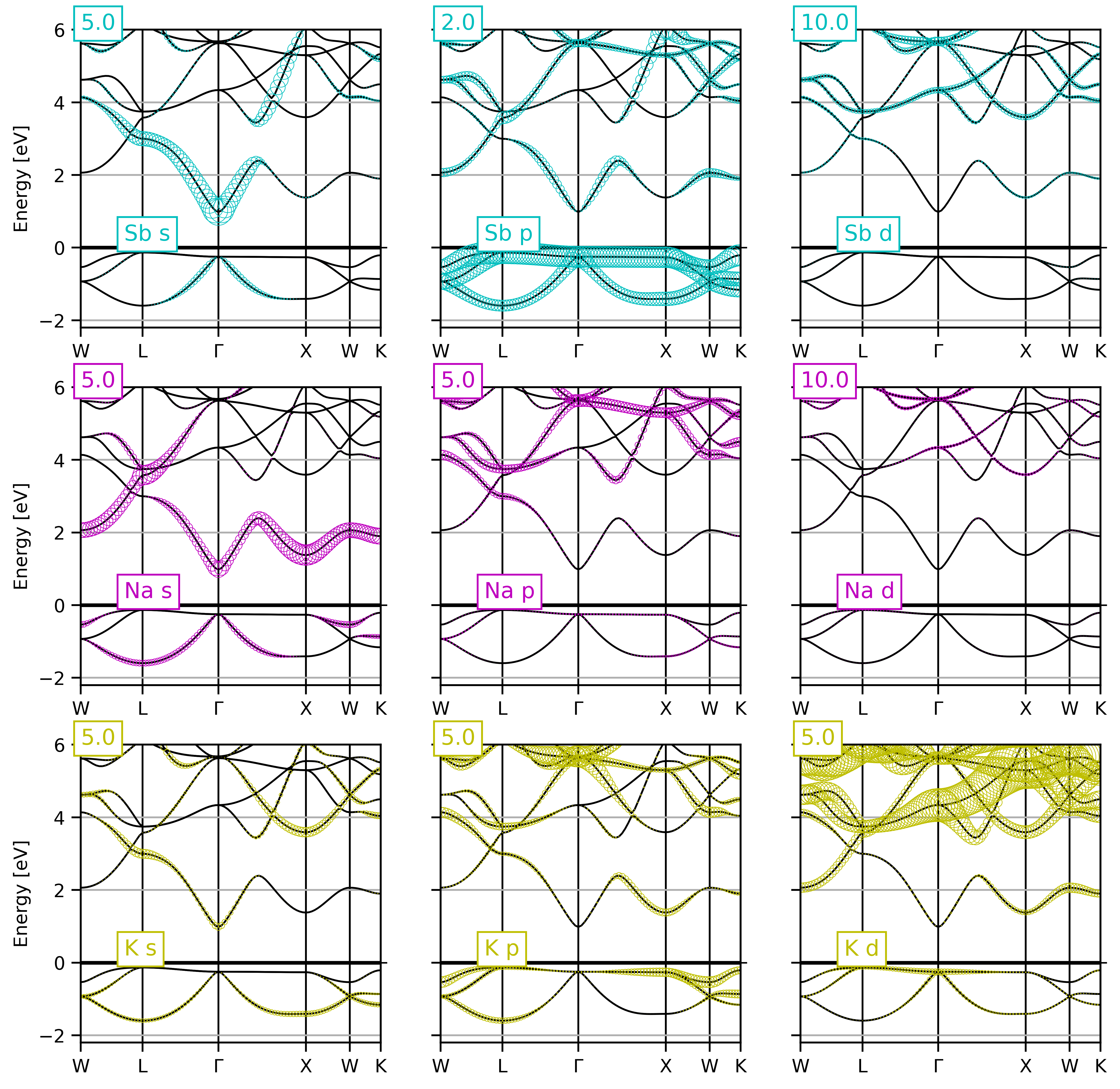}
    \caption{\label{fig:nak2sbweights}QP-corrected band-structure plots with the atom-projected character of the valence and conduction bands of cubic \cubcrys, quantified by the size of the colored circles. The magnification factor applied to the output data to improve visualization is reported in the top left corner of each plot. The energy scale is offset with respect to the Fermi energy highlighted by a horizontal black line. 
    } 
\end{figure*}

\begin{figure*}[h!]
    \centering  \includegraphics[width=\textwidth]{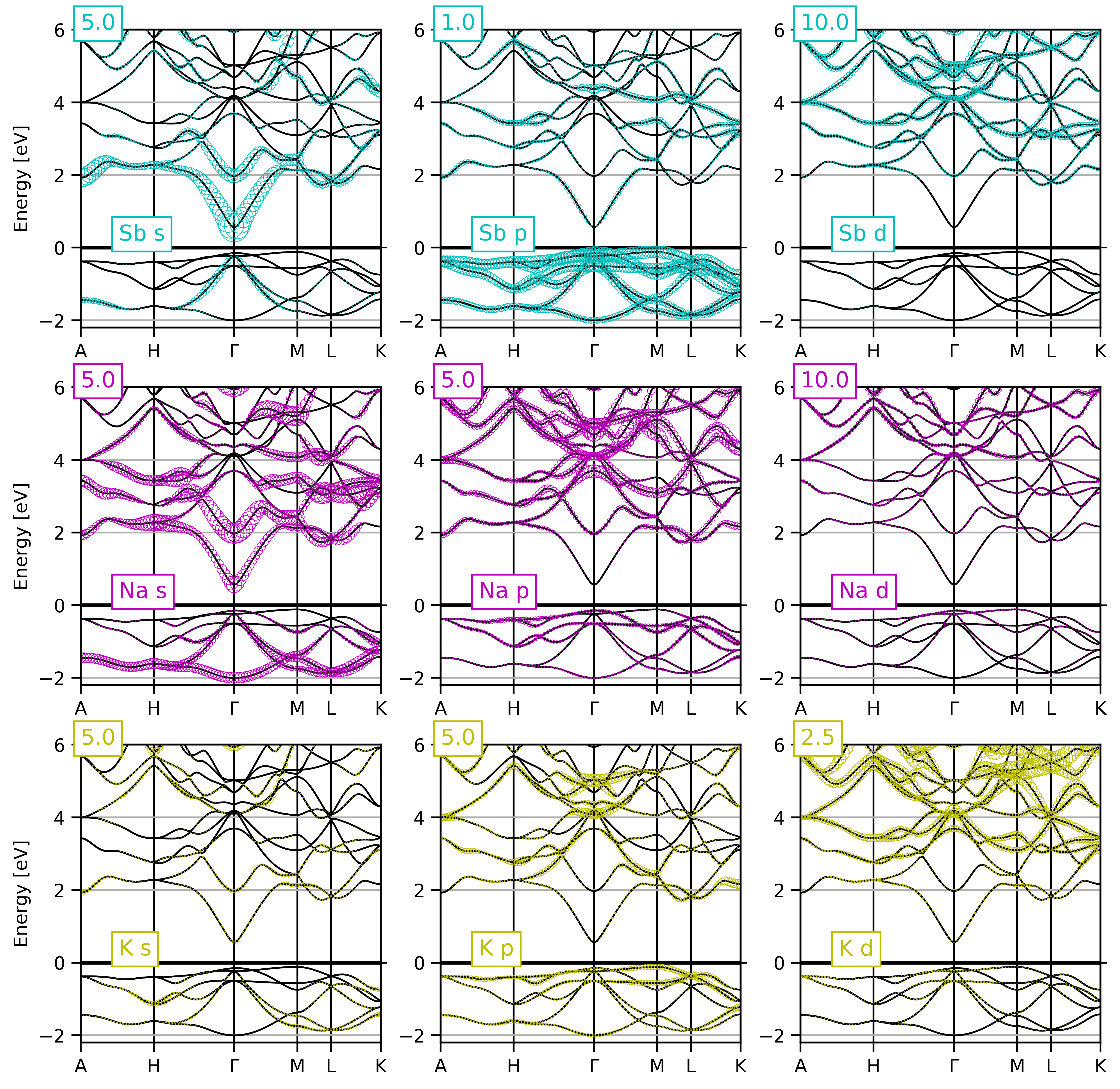}
    \caption{\label{fig:na2ksbweights}QP-corrected band-structure plots with the atom-projected character of the valence and conduction bands of hexagonal \hexcrys, quantified by the size of the colored circles. The magnification factor applied to the output data to improve visualization is reported in the top left corner of each plot. The energy scale is offset with respect to the Fermi energy highlighted by a horizontal black line.
    } 
\end{figure*}

To gain a deeper understanding of the electronic structure of the considered materials, we investigate the atomic-orbital contributions to the band strcturess as reported in Fig.~\ref{fig:nak2sbweights} for \cubcrys~and in Fig.~\ref{fig:na2ksbweights} for \hexcrys.
Sb $s$-orbitals dominate the CBm at $\Gamma$ and its surrounding in both crystals, in agreement with the results obtained for hexagonal \cubcrys~and cubic \hexcrys~\cite{amador}, and, more generally, for MAAs~\cite{Cocchi_2019, csk2sb_PBE, Cs_cocchi-2019, Sassnick_2021}. The uppermost valence bands have an Sb $p$-character, again in line with the known characteristics of other Cs- and Na-based MAAs~\cite{amador,Cs_cocchi-2019, csk2sb_PBE,Cocchi_2019, Sassnick_2021}. The Sb $d$-states contribute at higher energy in the conduction region, above the lowest band. It should be noticed that their weights are magnified by a factor of 10 in both Fig.~\ref{fig:nak2sbweights} and Fig.~\ref{fig:na2ksbweights}.

The CBm of both crystals is characterized by hybridized Na-Sb $s$-states (Figs.~\ref{fig:nak2sbweights} and \ref{fig:na2ksbweights}, middle panels). In cubic \cubcrys, they dominate the lowest conduction band along the $\Gamma\rightarrow$ X $\rightarrow$ W $\rightarrow$ K path (Fig.~\ref{fig:nak2sbweights}), and in hexagonal \hexcrys, they participate in most of the unoccupied bands shown in Fig.~\ref{fig:na2ksbweights}. In both crystals, Na $s$-states contribute also to the lowest state in the valence-band manifold.
Contributions from Na $p$-orbitals are spread in the conduction region of \hexcrys~above the CBm (Fig.~\ref{fig:na2ksbweights}), while in \cubcrys, they are mostly focused on bands a few eV above the gap (Fig.~\ref{fig:nak2sbweights}). In addition, they partially contribute to the valence states in \hexcrys, while they are absent in the uppermost occupied region of \cubcrys.
Finally, Na $d$-states are almost absent from the region visualized in Figs.~\ref{fig:nak2sbweights} and \ref{fig:na2ksbweights}. Weak contributions to the conduction region of both crystals are visible only through a 10$\times$ magnification.

Potassium $s$-orbitals do not significantly contribute to the electronic bands visualized in Figs.~\ref{fig:nak2sbweights} and \ref{fig:na2ksbweights}, bottom panels.
Instead, $p$-states participate in the conduction states $\sim$4~eV above the CBm in both materials.
The contribution of K $d$-orbitals is substantial in the conduction regions of both crystals from about 3~eV above the CBm, as indicated by the weights of these states magnified only by a factor of 5 in Figs.~\ref{fig:nak2sbweights} and \ref{fig:na2ksbweights}.
These results are overall in good agreement with those reported in Ref.~\cite{amador} for cubic \hexcrys~and hexagonal \cubcrys. This matching confirms that the character of the electronic bands is mainly dominated by the atomic composition of the materials rather than by their stoichiometry and crystal structure.

\subsection{\label{subsec:opticalres}Optical properties and excitons}%

\begin{figure}[h!]
    \centering
      \includegraphics[width=0.48\textwidth]{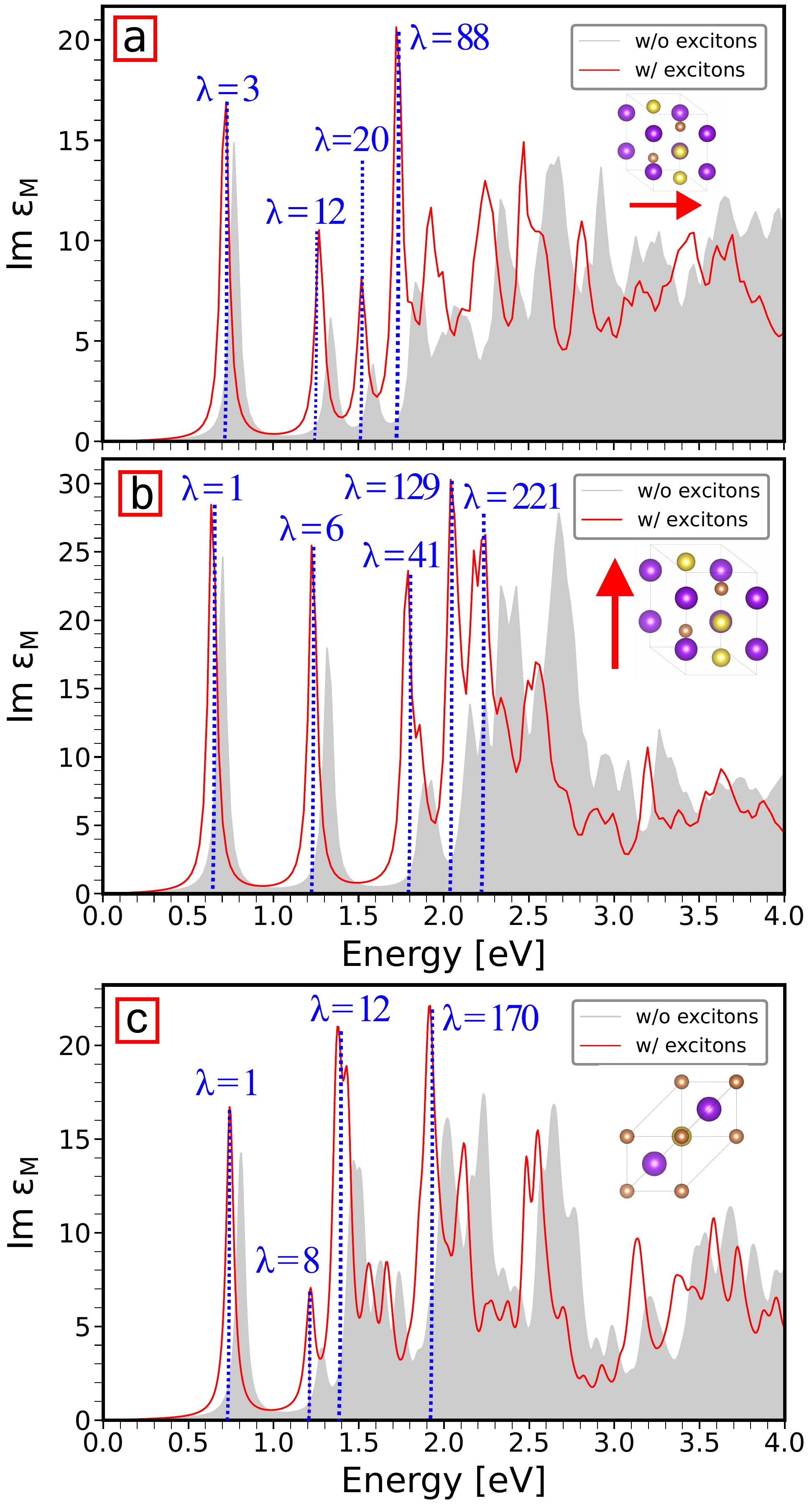}
\caption{\label{fig:BSE} Optical absorption spectra of hexagonal \hexcrys, including a) the in-plane and b) out-of-plane components of the imaginary part of the dielectric function, see schemes in the insets, and c) cubic \cubcrys~sketched in the insets. Spectra including excitonic effects as computed from the solution of the BSE are shown by red lines, while their IQPA counterpart without excitonic effects are indicated by the gray shaded area. Selected excitons are marked by dotted blue bars and labeled according to their $\lambda$ index in the solution of the BSE (Eq.~\eqref{eqn:BSE}). 
} 
\end{figure}

We now turn to the analysis of the optical properties of hexagonal \hexcrys~and cubic \cubcrys. In Fig.~\ref{fig:BSE}, we show their absorption spectra computed by solving the BSE [Eq.~\eqref{eqn:BSE}] by full diagonalization on top of the QP electronic structure to have direct access to eigenvalues $E^\lambda$ and eigenvectors $A^\lambda$ for each excitation. We compare the BSE spectra with those computed in the independent QP approximation (IQPA), where the BSE Hamiltonian includes only the diagonal term and electron-hole interactions are neglected: $\hat{H}^\text{BSE} = \hat{H}^{\rm diag}$. To account for the structural anisotropy of hexagonal \hexcrys, we plot in Figs.~\ref{fig:BSE}a,b both the in-plane and out-of-plane components of $\Im \varepsilon_M$, while for cubic \cubcrys, the absorption spectrum is given by a single optical component (Fig.~\ref{fig:BSE}c).

\begin{figure*}[h!]
  \centering
 \includegraphics[width=1\textwidth]{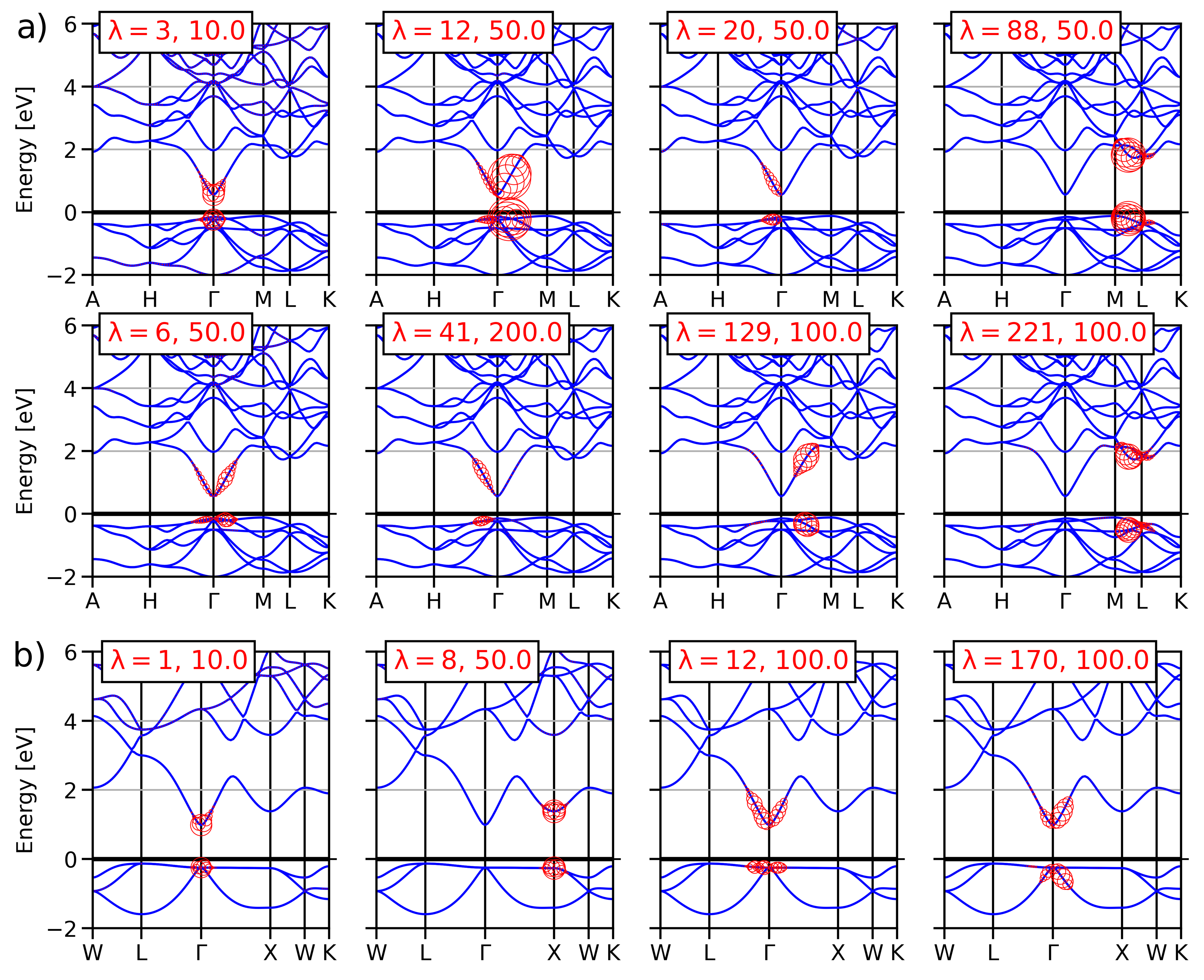}
  \caption{Exciton weights for selected excitations of a) hexagonal \hexcrys~and b) cubic \cubcrys. In each plot, the index of the excitation $\lambda$ and the magnification factor adopted for the colored circles indicating $w_h$ and $w_e$ to enhance visualization are indicated in the legend.} 
  \label{fig:exciton}
\end{figure*}

At a glance, the two components of the spectrum of hexagonal \hexcrys~exhibit similar features, consisting of sharp peaks at the onset located in the near-infrared region, see Figs.~\ref{fig:BSE}a and b. Additional narrow maxima appear at higher energies at the lower boundary of the visible range. The absorption continuum starts around 2~eV regardless of polarization. 
The first excitation, $\lambda=1$, is optically active in the out-of-plane direction and appears at an energy of \SI{0.64}{eV}, while the lowest-energy peak in the in-plane component of the spectrum corresponds to the third excitation ($\lambda=3$) and is found at $\SI{0.72}{\eV}$. Both features have a negligible excitonic character as indicated by the very similar energy, oscillator strength, and spectral shape of their counterparts in the IQPA. The binding energy of both $\lambda=1$ and $\lambda=3$ is on the order of 50~meV, \textit{i.e.}, half the value obtained for CsK$_2$Sb~\cite{Cocchi_2019}, hexagonal \cubcrys, and cubic \hexcrys~\cite{amador} at the same level of theory. They both stem from transitions between the uppermost valence band at $\Gamma$ and the CBm, see Fig.~\ref{fig:exciton}a ($\lambda=3$). The different polarization of $\lambda=1$ and $\lambda=3$ can be understood considering that the initial states of both transitions belong to a manifold of Sb $p$-states (see Fig.~\ref{fig:na2ksbweights}) featuring the typical spatial anisotropy of these atomic orbitals. The final state has an Sb $s$-character partially hybridized with Na $s$-orbital contributions.

The second absorption peak in the out-of-plane component of the spectrum of hexagonal \hexcrys~at  $\SI{1.23}{\eV}$ (Fig.~\ref{fig:BSE}b) corresponds to the sixth excitation ($\lambda=6$), which is slightly weaker than $\lambda=1$. In the in-plane component, the second maximum is given by $\lambda=12$ with an intensity of about two-thirds compared to $\lambda=3$ (Fig.~\ref{fig:BSE}a). Both resonances stem from vertical transitions distributed in the vicinity of $\Gamma$ toward the high-symmetry point M (see Fig.~\ref{fig:exciton}a). 
The nature of the initial and final states involved in the corresponding transitions are again Sb $p$ and $s$, respectively (see Fig.~\ref{fig:na2ksbweights}), as for the lowest-energy peaks discussed above. 

The third peak in the spectra of hexagonal \hexcrys~is found at 1.52~eV in the in-plane component (Fig.~\ref{fig:BSE}a) and corresponds to $\lambda=20$. In the out-of-plane direction, the third maximum is $\lambda=41$ at 1.78~eV (Fig.~\ref{fig:BSE}b). Both excitations exhibit a more pronounced excitonic character than the lower-energy peaks: their binding energy approaches 60~meV and their oscillator strength is enhanced by the electron-hole interactions included in the BSE solution compared to the IQPA. They both come mainly from transitions between $\Gamma$ and H (Fig.~\ref{fig:exciton}a) and the involved electronic states at the top of the valence band and at the bottom of the conduction region have Sb $p$-character and a hybridized Sb-Na $s$-nature, respectively (Fig.~\ref{fig:na2ksbweights}).

The continuum onset in the in-plane component of the spectrum of hexagonal \hexcrys~starts with $\lambda =88$, corresponding to the brightest peak appearing in Fig.~\ref{fig:BSE}a. This excitation is in the visible region at $\SI{1.93}{\eV}$ and displays a marked excitonic character testified by its stronger intensity and a 67~meV red-shift compared to its IQPA counterpart. It receives contributions from vertical transitions between M and L and, as such, it has a different character compared to the lower-energy states dominated by transitions around $\Gamma$ (see Fig.~\ref{fig:exciton}a). The initial states contributing to $\lambda =88$ have still predominantly an Sb $p$-character, although the contribution from K $p$-orbitals is larger than at the zone center. On the other hand, the conduction states involved in these transitions have a hybrid Sb-Na $s$-character with non-negligible contributions from K-Na $p$-orbitals.

Excitation $\lambda =129$ marks the continuum onset in the out-of-plane direction of the spectrum of hexagonal \hexcrys~and it is the most intense peak visualized in Fig.~\ref{fig:BSE}b. In contrast to $\lambda =88$, it comes from transitions between $\Gamma$ and M (Fig.~\ref{fig:exciton}), carrying predominantly Sb $p$- and $s$-character in the initial and final states of the transition, respectively (Fig.~\ref{fig:na2ksbweights}). The first bright excitation between M and L in the out-of-plane polarization direction is $\lambda =221$ at 2.24~eV and it appears even more intense than $\lambda =88$. Unfortunately, being located in the absorption continuum, it is not straightforward to evaluate its binding energy with respect to its IQPA counterpart.

The absorption spectrum of cubic \cubcrys~(Fig.~\ref{fig:BSE}c) shares similarities with hexagonal \hexcrys~but also exhibits distinct features. The first bright peak corresponds to the first excitation, $\lambda=1$, and appears at $\SI{0.64}{\eV}$. It does not display a pronounced excitonic character as its shape, intensity, and energy are very similar to its IQPA counterpart and its binding energy is $\SI{65}{\milli\eV}$. Due to the high symmetry of cubic \cubcrys, this state is triply-degenerate and stems from transitions between the topmost valence states at $\Gamma$ and the CBm, see Fig.~\ref{fig:exciton}. As shown in Fig.~\ref{fig:nak2sbweights}, the initial and final states of these transitions possess Sb $p$- and $s$-character, respectively. 

The second bright excitation corresponds to $\lambda = 8$ at $1.20$~eV which is also triply-degenerate. Its oscillator strength is about half the one of $\lambda=1$ and its binding energy is \SI{67}{\milli\eV}. It stems from transitions at X (Fig.~\ref{fig:exciton}b) where the valence states are predominantly K $p$-orbitals whereas in the conduction they have a hybrid K $d$- and Na $s$-character (Fig.~\ref{fig:nak2sbweights}). 
The third excitation at $\lambda = 12$ exhibits a higher intensity than the first one and a binding energy of 78~meV. This transition, situated just below the lower edge of the visible spectrum ($\SI{1.40}{\eV}$), originates from electronic states distributed across the upper valence band and the vicinity of the CBM near $\Gamma$, exhibiting a significant Sb $p$- and $s$-character, respectively,  see Fig.~\ref{fig:nak2sbweights}.
The spectral region between 1.50~eV and 2.00~eV hosts a large density of excitations with $\lambda = 170$ at $\SI{1.91}{\eV}$ representing the sharpest peak in Fig.~\ref{fig:BSE}c. Its IQPA counterpart can be recognized in the corresponding spectrum \SI{74}{\milli\eV} above in energy. The relatively pronounced excitonic effects exhibited by this excitation include also the gain of oscillator strength due to electron-hole correlations. Remarkably, its character is rather similar to $\lambda = 12$, being dominated by transitions around $\Gamma$ although slightly oriented toward X and involving a deeper valence band. Both the initial and final states are primarily composed of Sb $p$- and $s$-orbitals, respectively (Fig.~\ref{fig:nak2sbweights}).

Before concluding, it is instructive to discuss the results presented in this section with those obtained for cubic \hexcrys~and hexagonal \cubcrys~in Ref.~\cite{amador}. We recall that the latter crystals are the experimentally known phases of the materials. 
Consistent with the lower values of the QP gaps, the absorption onsets of hexagonal \hexcrys~and cubic \cubcrys~appear at lower energies compared to those of their cubic and hexagonal counterparts, respectively. Exciton binding energies are lower in the computationally predicted polymorphs considered here than in the experimental phases discussed in Ref.~\cite{amador}. This characteristic is compatible with the lower QP gap and, thus, the slightly larger values of static screening entering the screened Coulomb interaction $W = \epsilon^{-1}v$: in hexagonal \hexcrys, the in-plane and out-of-plane components of $\Re\epsilon(\omega=0)$ are 9.14 and 11.48, respectively, while for cubic \cubcrys, $\Re\epsilon(\omega=0)=9.71$. 
Another noteworthy point is that in both hexagonal \hexcrys~and cubic \cubcrys, the lowest-energy excitation is optically active. This is not the case for hexagonal \cubcrys, where the first excitation is dark~\cite{amador}. 


\section{\label{sec:conclusion}Summary and conclusion}%
In summary, we presented an \textit{ab initio} study based on DFT and MBPT ($G_0W_0$ and BSE) of the electronic and optical properties of hexagonal \hexcrys~and cubic \cubcrys, two computationally predicted phases of established multi-alkali antimonide materials used as photocathodes for particle accelerators. Both crystals exhibit an indirect band-gap, thus differing from the experimentally known counterparts, cubic \hexcrys~and hexagonal \cubcrys, which are both direct band-gap semiconductors~\cite{amador}. However, the direct gaps at $\Gamma$, on the order of 0.7~eV in hexagonal \hexcrys~and 0.8~eV in cubic \cubcrys, are very close to the fundamental ones.   
Their optical absorption spectra are characterized by intense and distinct absorption peaks in the near-infrared region, where the lowest-energy excitations appear. Excitonic effects are negligible in these low-energy excitations which carry clear single-particle signatures testified by their spectral shape reproduced in the independent quasi-particle approximation and binding energies of the order of 50~meV. On the other hand, electron-hole correlations play a bigger role in the higher-energy excitations appearing in the visible region. In hexagonal \hexcrys, where the optical absorption is anisotropic, the lowest-energy peaks stem from the transitions at $\Gamma$, where the involved electronic states are dominated by Sb states with $p$- and $s$-character in the valence and conduction region, respectively. The higher energy excitations, instead, originate from transitions between the high-symmetry points M and L, involving hybrid Sb-K occupied $p$-orbitals and hybrid unoccupied Sb-Na $s$-states.
In the spectrum of cubic \cubcrys, the most intense excitations exhibit degeneracy due to the symmetry of this material. They are mostly due to transitions around $\Gamma$, involving Sb $p$- and $s$-states, except for a weaker excitation at 1.2~eV stemming from transitions among states with potassium $p$-character in the valence and $d$-character in the conduction region.
This composition is shared with the experimentally stable polymorphs hexagonal \cubcrys~and cubic \hexcrys, although in these two compounds, the first excitations are slightly higher and possess weaker oscillator strength in line with the trends obtained for the fundamental gaps~\cite{amador}. 

The electronic and optical properties of hexagonal \hexcrys~and cubic \cubcrys~make them suitable candidates for photocathode applications. In particular, the optical gap in the near-infrared region and the weak binding energy of their excitations align with the current requirement for efficient vacuum electron sources in particle accelerators~\cite{ADVANCES_ESOURCE+MUSUMECI2018209}. While the existence of these compounds has not been experimentally proven yet, we are confident that this work may stimulate research in this direction. Most importantly, our results indicate that the presence of these phases in polymorphic samples is not detrimental to their photocathode performance.

\section*{Acknowledgments}
This work was funded by the German Research Foundation (DFG), Project No. 490940284, DAAD (Program RISE), the German Federal Ministry of Education and Research (Professorinnenprogramm III), and the State of Lower Saxony (Professorinnen f\"ur Niedersachsen). Computational resources were provided by the HPC cluster ROSA at the University of Oldenburg, funded by the DFG (project number INST 184/225-1 FUGG) and by the Ministry of Science and Culture of the Lower Saxony State.

\section*{Data availability statement}
The data that support the findings of this article are openly available:~\url{https://doi.org/10.5281/zenodo.14100233}. 

\selectlanguage{english}

%

\end{document}